
\magnification=1200
\def\Quadrat#1#2{{\vcenter{\hrule height #2
  \hbox{\vrule width #2 height #1 \kern#1
    \vrule width #2}
  \hrule height #2}}}
\def\dAlembert{\mathop{\kern 1pt\hbox{$\Quadrat{6pt}{0.4pt}$} \kern1pt}}

\def\d{\partial}
\def\ref#1{\lbrack#1\rbrack}
\def\next{\hfil\break\noindent}

\font\title=cmbx12
{\title
\centerline{Blow-Up of Test Fields Near Cauchy Horizons}}
\vskip 2cm\noindent
J. Rauch\footnote*{Partially supported by NSF grant DMS 8601783 and Office
of Naval Research grant ONR NO 014 92 J 1245}
\next
Department of Mathematics
\next
University of Michigan
\next
Ann Arbor MI 48109
\next
U.S.A.
\vskip 10pt\noindent
and
\vskip 10pt\noindent
A. D. Rendall
\next
Max-Planck-Institut f\"ur Astrophysik
\next
Karl-Schwarzschild-Str. 1
\next
8046 Garching bei M\"unchen
\next
Germany
\vskip 5cm
\noindent
{\bf Abstract}

The behaviour of test fields near a compact Cauchy horizon is
investigated. It is shown that solutions of nonlinear wave
equations on Taub spacetime with generic initial data cannot be
continued smoothly to both extensions of the spacetime through the
Cauchy horizon. This is proved using an energy method. Similar results
are obtained for the spacetimes of Moncrief containing a compact Cauchy
horizon and for more general matter models.
\vskip 20pt

\vfil\eject
{\baselineskip=20pt

\noindent
{\bf 1.Introduction}

The cosmic censorship hypothesis is one of the most important open
problems in general relativity. Penrose\ref1\ has presented a
plausibility argument in favour of strong cosmic censorship and it
is of interest to ask how this could be turned into a rigorous proof.
The basic idea of the argument is that if a spacetime contains a
Cauchy horizon then a generic small perturbation of the initial data on
a partial Cauchy surface should lead to an accumulation of energy
or infinite blueshift and that as a result the Cauchy horizon would
be replaced by a curvature singularity. There are various ways in
which one could attempt to capture an effect of this kind
mathematically. One is to attempt to show the blow-up of solutions of
the Einstein equations along appropriate curves using some kind of
geometrical optics approximation \ref2. Another
approach is to attempt to use energy inequalities. These are usually
employed in the proofs of existence theorems to show that certain
quantities remain bounded. However such
inequalities can also be used to prove blow-up in
certain circumstances. In this paper a procedure of this kind will
be used successfully, in a situation simpler than that
of the above discussion.

In order to make progress we make some simplifying assumptions. Instead
of the Einstein equations themselves, we study
test fields i.e. solutions of various hyperbolic
equations on a {\it fixed spacetime background}. The goal is to prove that
generic solutions of these equations blow up near a Cauchy horizon of
the background spacetime. The relevance of such a result to the original
problem is twofold. On the one hand one would expect from a physical
point of view that if a spacetime is such as to cause matter fields
propagating through it to become singular this would lead in a more
realistic description, where the matter fields are coupled to the
Einstein equations, to a spacetime singularity as well. On the other
hand, the study of test fields may lead to the discovery of
mathematical ideas which will also be important when it comes to
solving the full problem. It follows from the latter point that theorems
concerning test fields will be particularly interesting if they use
techniques which have a good chance of being generalised. An advantage
of the energy method is that it has a wide potential range of
applicability. Its disadvantage is that the information obtained
concerns global integral quantities; it is difficult to localise
singularities which occur.

The basic idea of the energy method will now be recalled. Suppose
that a spacetime has been chosen along with a kind of test field
whose propagation on this spacetime is to be studied. Let
$T^{\alpha\beta}$ denote the energy-momentum tensor and let $S_t$
be a leaf of a foliation of part of spacetime by compact spacelike
hypersurfaces. The leaves are indexed by real numbers $t$ lying
in some interval. The assumption of compactness is not essential
but is convenient and also sufficient for most of the applications in
this paper. Let $G$ be a compact region bounded by the hypersurfaces
$S_t$ and $S_{t^\prime}$. Denote the normal vector field to the
foliation by $n^\alpha$. Then Stokes theorem and the divergence-free
nature of the energy-momentum tensor imply the identity
$$\int_{S_{t^\prime}}T_{\alpha\beta}n^\alpha p^\beta
=\int_{S_t}T_{\alpha\beta}n^\alpha p^\beta
+\int_G T^{\alpha\beta}p_{\alpha;\beta}\eqno(1)$$
for any smooth vector field $p^\alpha$. Of course for the Einstein
equations themselves this identity is not useful. The gravitational
field does not have an energy-momentum tensor. In that case, however,
it is possible to envisage replacing $T^{\alpha\beta}$ by the
Bel-Robinson tensor, a strategy which was used to good effect in
\ref3.

In the following section it will be shown how the above strategy can
be used to obtain information about scalar wave equations on the
Taub spacetime. Section 3 contains generalisations of this to a class
of spacetimes with compact Cauchy horizons studied by Moncrief, to other
kinds of matter models and to certain spacetimes with non-compact
Cauchy horizons.

\vskip .5cm
\noindent
{\bf 2.Scalar wave equations on the Taub spacetime}

This section is concerned with nonlinear wave equations of the form
$$\dAlembert u=m^2 u+\lambda u^3\eqno(2)$$
where $m$ and $\lambda$ are non-negative real numbers. They will be
studied on the spacetime with metric
$$\eqalign{ds^2=-U^{-1}dt^2+(2l)^2 U(d\psi+cos&\theta d\phi)^2       \cr
&+(t^2+l^2)(d\theta^2+\sin^2\theta d\phi^2),}\eqno(3)$$
The notation here coincides with that used by Hawking and Ellis\ \ref4,
p.170. The function $U$ depends only on $t$, and $l$ is a positive
constant. Equation (3) defines a non-degenerate metric so long as
$U$ is positive. The Taub spacetime is given by the particular function
$U(t)=(l^2-2mt-t^2)/(t^2+l^2)$ and is a solution of the vacuum Einstein
equations. All that is required for our blowup theorem is that $U$ is
strictly positive on an interval $(t_-,t_+)$ and is zero
with nonvanishing derivatives at the
endpoints. The Taub spacetime is defined on the manifold
$M=S^3\times (t_-,t_+)$. It is well known (see \ref4) that $t=t_+$ is
only a coordinate singularity and that the Taub spacetime can be
extended through this hypersurface in two distinct ways. This gives rise
to the Taub-NUT spacetime for which $t=t_+$ is a Cauchy horizon. Let
$M_1$ denote the manifold where one extension is defined and $M_2$
the other. If
$$\psi^\prime:=\psi+(1/2l)\int U^{-1}(t) dt,\eqno(4)$$
then $(t,\psi^\prime,\theta,\phi)$ are regular coordinates on $M_1$
(apart from the trivial singularities due to the use of polar
coordinates). Similarly
$\psi^{\prime\prime}=\psi-(1/2l)\int U^{-1}(t)dt$ defines
together with $t$, $\theta$ and $\phi$ a regular coordinate system on
$M_2$. The explicit form of the linear wave equation in Taub space is
$$\eqalign{
&-Uu_{tt}-U_tu_t+\lbrack(t^2+l^2)^{-1}\cot^2\theta+(4l^2U)^{-1}\rbrack
u_{\psi\psi}+(t^2+l^2)^{-1}u_{\theta\theta}               \cr
&+(t^2+l^2)^{-1}\csc^2\theta u_{\phi\phi}-2(t^2+l^2)^{-1}\csc\theta\cot
\theta u_{\theta\phi}-(t^2+l^2)^{-1}u_\theta=0.}\eqno(5)$$
Consider for a moment the possibility that $u$ only depends on $t$.
Then (5) reduces to
$$Uu_{tt}+U_tu_t=0,\eqno(6)$$
with the explicit solution
$$u=C_1+C_2\int U^{-1}(t) dt.\eqno(7)$$
Hence there are special solutions of (5) which blow up near the Cauchy
horizon i.e. as $t\to t_+$. For the linear equation (5) this is
enough to show that the generic solution blows up (since it is always
possible to add $\epsilon$ times the data for this singular solution
to any data whose corresponding solution does not blow up) but for a
non-linear equation this no longer works. It is necessary to use
methods which are more flexible.

The energy identity (1) will now be applied to equation (2) on the Taub
spacetime. The energy-momentum tensor associated to the equation (2)
is given by
$$T^{\alpha\beta}=\nabla^\alpha u\nabla^\beta u-(1/2)g^{\alpha\beta}
(\nabla_\gamma u\nabla_\delta u g^{\gamma\delta}+m^2u^2+(1/2)\lambda
u^4).\eqno(8)$$
That it is divergence free follows directly from (2).
The foliation chosen is that given by the hypersurfaces of
constant $t$. As the vector field $p^\alpha$ choose $\d/\d\psi$. This
extends smoothly to both $M_1$ and $M_2$ since it is equal to
$\d/\d\psi^\prime$ and $\d/\d\psi^{\prime\prime}$ in the regions
where the relevant coordinate patches overlap. This is a Killing vector
and so the volume contribution in (1) vanishes. It follows that
$\int_{S_t}T_{\alpha\beta}n^\alpha p^\beta$ is independent of $t$. To
see what this means the integral is evaluated explicitly.
The normal vector $n^\alpha$ to the foliation is given in the
coordinates used in (3) by $U^{1/2}\d/\d t$. Hence
$$T_{\alpha\beta}n^\alpha p^\beta=U^{1/2}u_t u_\psi.\eqno(9)$$
Next note that the expression $\sin\theta d\psi d\theta d\phi$ defines a
smooth volume element on each leaf of the foliation which extends
regularly to $M_1$ and $M_2$. Also the volume element induced on $S_t$
by the given metric is $2lU^{1/2}(t^2+l^2)\sin\theta d\psi d\theta
d\phi$. Thus the integral can be written in the form
$$\int_{S_t}T_{\alpha\beta}n^\alpha p^\beta=l^2U^2\int \left(
{\d u\over\d t}+{1\over 2lU}{\d u\over\d\psi}\right)^2
-\left({\d u\over\d t}
-{1\over 2lU}{\d u\over\d\psi}\right)^2 dV,
\eqno(10)$$
where $dV=(t^2+l^2)\sin\theta d\psi d\theta d\phi$.
Now in the overlap of the coordinate patches the vector field
$\d/\d t-(2lU)^{-1}\d/\d\psi$ on $M$ corresponds to the regular vector
field $\d/\d t$ on $M_1$ and similarly $\d/\d t+(2lU)^{-1}\d/\d\psi$
corresponds to $\d/\d t$ on $M_2$. It follows that if a solution $u$
of (2) had $C^1$ extensions to both $M_1$ and $M_2$ the right hand
side of (10) would tend to zero as $t\to t_+$. But we have already seen
that this expression is constant. Thus this extendibility property can
only hold for very special initial data, proving the following
theorem.

\vskip 10pt\noindent
{\bf Theorem 1} Let $t_0$ be a number in the interval $(t_-,t_+)$ and
suppose that initial data $(u,u_t)$ for equation (2) are given on
the spacelike hypersurface $t=t_0$ in Taub spacetime. Then if
$$\int_{S_{t_0}} u_{\psi}u_t \sin\theta
d\psi d\theta d\phi\ne0\eqno(11)$$
the corresponding solution of (2) cannot extend in a $C^1$ manner to
both $M_1$ and $M_2$.

\vskip 10pt\noindent
{\bf Remarks} 1. The condition (11) is generic in the sense that it is
fulfilled for an open dense set of initial data in the uniform topology.

\noindent
2. The particular form of the nonlinearity in (2) was chosen because it
ensures that for given initial data there exists a corresponding
solution globally on $M$\ref5. The theorem would still hold for a
more general nonlinearity but in that case blow-up could in principle
occur even before the Cauchy horizon.

\vskip .5cm
\noindent
{\bf 3.Further examples}

The existence of a large class of solutions of the vacuum Einstein
equations containing compact Cauchy surfaces has been shown by Moncrief
\ref6. These are somewhat similar to the Taub-NUT spacetime but differ
by the fact that they have partial Cauchy surfaces diffeomorphic to
$K\times S^1$, where $K$ is a compact surface, instead of $S^3$. The
form of the metric is
$$ds^2=e^{-2\phi}\lbrack -N^2dt^2+g_{ab}dx^adx^b\rbrack+t^2e^{2\phi}
(dx^3+\beta_adx^a)^2.\eqno(12)$$
Here Latin indices take the values 1 and 2 and $x^3$ is the coordinate
on $S^1$. The metric coefficients do not depend on $x^3$.
The function $N$ is assumed  to vanish nowhere. This form of
the metric becomes singular at $t=0$. However, as in the case of the
Taub spacetime, it is possible to extend the metric through the
hypersurface $t=0$ in two different ways to manifolds $M_1$ and $M_2$
provided a certain additional condition is satisfied. This is seen by
introducing new coordinates $x^{3\prime}=x^3-\log t$ and
$x^{3\prime\prime}=x^3+\log t$ respectively and $t^\prime=t^2$.
The condition for the existence of the extensions is then that
$(N^2-e^{4\phi})/t^\prime$ is a smooth function of the new coordinates.
The spacetimes constructed by Moncrief are all analytic. This condition
plays no role in the present considerations, where finite
differentiability of the metric (12) is sufficient. It will be shown
that the above arguments concerning solutions of (2) on Taub space can
easily be adapted to this case.

The equivalent of (6) in this case is
$tu_{tt}+u_t=0$ (which can be read off from the formulae in \ref5)
with solution $u=C_1+C_2\log t$. For the application
of the energy identity choose the foliation $t=$const. and let the
vector field $p^\alpha$ be given by $\d/\d x^3$. Since $\d/\d x^3$ is
a Killing vector there is no volume contribution in (1). The normal
vector $n^\alpha$ to the foliation is $e^{\phi}N^{-1}\d/\d t$. The
volume element induced on $S_t$ by the spacetime metric is
$$te^\phi\lbrack g_{11}g_{22}-g_{12}^2+2g_{12}\beta_1\beta_2
-g_{11}\beta_2^2-g_{22}\beta_1^2\rbrack^{1/2} dx^1dx^2dx^3.\eqno(13)$$
Now the metric
$$e^{-2\phi}\lbrack -N^2dt^2+g_{ab}dx^adx^b\rbrack+e^{2\phi}
(dx^3+\beta_adx^a)^2\eqno(14)$$
is regular even at $t=0$ without doing a coordinate change. Hence
the expression obtained from (13) by omitting the factor $t$ is a
regular volume element, call it $dV$. Hence
$$\int_{S_t} T_{\alpha\beta}n^\alpha p^\beta={1\over2}\int e^{-\phi}
N^{-1}t^4\left\lbrack\left({1\over t}{\d u\over\d t}
+{1\over t^2}{\d u\over \d x^3}\right)^2-
\left({1\over t}{\d u\over\d t}-
{1\over t^2}{\d u\over\d x^3}\right)^2\right\rbrack
dV$$
must tend to zero as $t\to 0$ if the solution $u$ has a $C^1$
extension to both $M_1$ and $M_2$. On the other hand the energy
identity shows that this expression is independent of time. Thus
we get the following analogue of Theorem 1.
\vskip 10pt
\noindent
{\bf Theorem 2} Let $t_0$ be a positive real number and suppose that
initial data $(u,u_t)$ are given for equation (2) on the spacelike
hypersurface $t=t_0$ in the spacetime defined by the metric (12) on
the manifold $K\times S^1\times (0,\infty)$, where $K$ is a compact
surface. Suppose furthermore that the condition given above for the
existence of extensions $M_1$ and $M_2$ of the spacetime through $t=0$
is satisfied. Then if
$$\int_{S_{t_0}} e^{-\phi}N^{-1} u_tu_{x^3} dV\ne0\eqno(15)$$
the corresponding solution of (2) cannot extend in a $C^1$ manner to
both $M_1$ and $M_2$.

\vskip .25cm
The fact that in this theorem the Cauchy horizon occurs in the past of
the initial hypersurface (if the direction of increasing $t$ is chosen
as the future direction) in contrast to the situation in Theorem 1 is
merely a matter of notational convenience. In \ref5\ Moncrief analyses
solutions of equation (2) on spacetimes of the form (12) with regard to
their behaviour near the Cauchy horizon. His results do not seem
to be directly comparable with those of the present paper.

Next we consider the possibility of applying the energy method to
other matter models. For simplicity only the case of Taub spacetime
will be written out but the arguments generalise to Moncrief's
spacetimes. The assumption that a solution only depends on $t$ has
no natural analogue for geometric objects more complicated than
scalars and so the cheap construction of solutions which blow up
on the Cauchy horizon in the scalar case is not available. However
the energy method still works, as will now be demonstrated. Denote
the vectors $\d/\d t+(2lU)^{-1}\d/\d\psi$ and $\d/\d t-(2lU)^{-1}
\d/\d\psi$ by $n_1^\alpha$ and $n_2^\alpha$ respectively. Then
$$n^\alpha=(1/2)(n_1^\alpha+n_2^\alpha)\eqno(16)$$
Hence
$$2T_{\alpha\beta}n^\alpha p^\beta=T_{\alpha\beta}n_1^\alpha p^\beta
+T_{\alpha\beta}n_2^\alpha p^\beta.\eqno(17)$$
If the solution extends smoothly to $M_1$ then the first term on
the right hand side of (16) is bounded while a smooth extension to
$M_2$ implies the boundedness of the second term. So the kind of
argument used twice already implies that the following theorem holds.
\vskip 10pt
\noindent
{\bf Theorem 3} Let $t_0$ be a number in the interval $(t_-,t_+)$
and suppose that initial data are given on the spacelike hypersurface
$t=t_0$ in Taub spacetime for some matter field with
energy-momentum tensor $T^{\alpha\beta}$. Then if
$$\int_{S_{t_0}}T_{\alpha\beta}n^\alpha p^\beta\ne0.\eqno(18)$$
the solution of the matter field equation corresponding to the
given initial data cannot extend to both $M_1$ and $M_2$ in a way
which would imply the continuous extendibility of the energy-momentum
tensor.

\vskip 10pt \noindent
{\bf Remark} The extendibility property is stated here in a
somewhat indirect way. For most matter models commonly used in general
relativity the energy-momentum tensor is expressible pointwise in
terms of the matter variables and their first covariant derivatives.
Thus $C^1$ extendibility of the matter variables implies continuous
extendibility of $T^{\alpha\beta}$.
\vskip .25cm
The only thing which varies from one matter model to another in
applying Theorem 3 is the explicit form of (18) in terms of the matter
variables. For a Maxwell field $F^{\alpha\beta}$, for instance, it takes
the form
$$\int_{S_{t_0}}(\csc^2\theta F_{03}F_{13}+F_{02}F_{12}-\csc\theta
\cot\theta F_{01}F_{13})\ne 0.\eqno(19)$$
The condition for a Yang-Mills field is identical to this except
that the expression on the left hand side of (19) is then
matrix-valued and the condition (18) is given by the vanishing of
its trace.

The methods used above can also be used to obtain information about
test fields on certain spacetimes containing non-compact Cauchy
horizons. Consider spacetimes of the Moncrief form where the condition
that the surface $K$ be compact is dropped. Then spacetimes will be
obtained which contain Cauchy surfaces diffeomorphic to $K\times S^1$.
If compactly supported data for some test field are given on a
hypersurface $t$=const then the support of the corresponding solution
on the globally hyperbolic region (i.e. the part of the spacetime before
the Cauchy horizon) will have compact closure in each of the extensions.
Also the intersection of this support with the other hypersurfaces of
constant $t$ are compact and so the identity (1) holds. Here it has
been assumed that the domain of dependence for the matter field
equations is limited by the light cones. A sufficient condition for
this to be true is that $T^{\alpha\beta}$ satisfies the dominant energy
condition. (For a discussion of this point see \ref4, p.94.) Thus the
same arguments as before apply. To sum up, suppose that condition that
$K$ be compact is removed from the hypotheses made on the spacetime in
Theorem 2 and a matter field is considered which satisfies the dominant
energy condition (for example the scalar field with energy-momentum
tensor given by (8)). Let compactly supported initial data for this
matter field be given on $t=t_0$ satisfying (18). Then the solution
which evolves from these initial data cannot have extensions to both
$M_1$ and $M_2$ which are regular enough to imply that the
energy-momentum has continuous extensions in both cases. Unfortunately
there are naturally occurring Cauchy horizons (such as that of the
Reissner-Nordstr\"om solution) which are not covered by this argument
since their Cauchy horizons have a different topology and compactly
supported initial data can be smeared over the whole horizon by the
time evolution.

\vskip .5cm
\noindent
{\bf References}
\next
1.Penrose, R. Singularities and time-asymmetry. In Hawking, S., Israel,
W. (eds.) {\it General relativity: an Einstein centenary survey.} Cambridge
University Press, Cambridge, 1979.
\next
2.Joly,J.L., Metivier, G., Rauch, J. Resonant one dimensional
geometric optics. {\it J. Funct. Anal.} {\bf 114}, 106-231 (1993).
\next
Joly,J.L., Metivier, G., Rauch, J. Remarques sur l'optique
g\'eometrique nonlin\'eaire multidimensionelle. S\'eminaire Equations
aux D\'eriv\'ees Partielles, Ecole Polytechnique, 1990-1991, expos\'e
no. 1;
\next
Schochet, S. Fast singular limits of hyperbolic equations. Preprint.
\next
3.Christodoulou, D., Klainerman, S. {\it The global nonlinear stability of
the Minkowski Space.} Princeton University Press, Princeton, 1993.
\next
4.Hawking, S.W., Ellis, G.F.R. {\it The large scale structure of space-time.}
Cambridge University Press, Cambridge, 1973.
\next
5.Moncrief, V. The asymptotic behaviour of nonlinear waves near a
cosmological Cauchy horizon. {J. Math. Phys.} {\bf 30}, 1760-1768 (1989).
\next
6.Moncrief, V. Neighbourhoods of Cauchy horizons in cosmological
spacetimes with one Killing field. {\it Ann. Phys.} {\bf 141}, 83-103 (1982).
}
\end